\UseRawInputEncoding
\documentclass[3p]{elsarticle}

\usepackage{lineno,hyperref}
\modulolinenumbers[5]

\journal{Journal of Computational Physics}

\makeatletter
\def\ps@pprintTitle{%
 \let\@oddhead\@empty
 \let\@evenhead\@empty
 \def\@oddfoot{\footnotesize\itshape
       Published in the \ifx\@journal\@empty Elsevier
       \else\@journal\fi\hfill} 
 \let\@evenfoot\@oddfoot}
\makeatother

\usepackage{lipsum}
\usepackage{amsfonts}
\usepackage{graphicx}
\usepackage{epstopdf}
\usepackage{soul}
\usepackage{float}
\usepackage{multirow}
\usepackage{amsbsy}
\usepackage{amssymb}
\usepackage{amsmath} 
\usepackage{algorithmic}
\usepackage{algorithm}
\usepackage{mathtools}
\usepackage{mathrsfs}
\usepackage{times}
\usepackage{enumerate}
\usepackage{color}
\usepackage{xspace}

\usepackage{subcaption}

\usepackage{tikz}
\usetikzlibrary{shapes,arrows}

\renewcommand{\b}{{\bf b}}
\renewcommand{\c}{{\bf c}}

\newcommand{\f}{{\bf f}}

\renewcommand{\v}{{\bf v}}

\newcommand{\x}{{\bf x}}
\newcommand{\y}{{\bf y}}
\newcommand{\z}{{\bf z}}

\newcommand{\B}{{\bf B}}

\newcommand{\I}{{\bf I}}

\newcommand{\N}{\mathcal{N}}  %
 
\newcommand{\Acal}{\mathcal{A}}

\newcommand{\Dcal}{\mathcal{D}}

\newcommand{\Lcal}{\mathcal{L}}
\newcommand{\Tcal}{\mathcal{T}}

\newcommand{\Scal}{\mathcal{S}}

\newcommand{\X}{{\mathbf{X}}}

\newcommand{\Mcal}{{\mathcal{M}}}

\newcommand{\balpha}{\boldsymbol{\alpha}}
\newcommand{\bpsi}{\boldsymbol{\psi}}

\newcommand{\btheta}{\boldsymbol{\theta}}

\newcommand{\bxi}{\boldsymbol{\xi}}
\newcommand{\bSigma}{\boldsymbol{\Sigma}}

\newcommand{\bmu}{\boldsymbol{\mu}}

\newcommand{\0}{{\bf 0}}

\newcommand{\ben}{\begin{enumerate}}
\newcommand{\een}{\end{enumerate}}

\newcommand{\argmin}{\operatornamewithlimits{argmin}}

\newcommand{\ie}{{i.e.,}\xspace}
\newcommand{\eg}{{e.g.,}\xspace}
\newcommand{\etc}{{etc.}\xspace}

\newcommand{\cmt}[1]{}

\newcommand{\RR}{\mathbb{R}}

\newcommand{\bPsi}{\boldsymbol{\Psi}}
\newcommand{\bXi}{\boldsymbol{\Xi}}
\newcommand{\btx}{\textbf{\textit{x}}}
\newcommand{\bty}{\textbf{\textit{y}}}
\newcommand{\btz}{\textbf{\textit{z}}}

\newcommand{\bupsilon}{\boldsymbol{\upsilon}} 

\newcommand{\bmaml}{{{B-MAML}}\xspace}
\newcommand{\pmaml}{{{P-MAML}}\xspace}
\newcommand{\emaml}{{{E-MAML}}\xspace}
\newcommand{\imaml}{{{I-MAML}}\xspace}
\newcommand{\maml}{{{MAML}}\xspace}
\newcommand{\ours}{{{BELTS}}\xspace}

\begin{document}

\begin{frontmatter}

\title{A Metalearning Approach for Physics-Informed Neural Networks (PINNs): Application to Parameterized PDEs}

\author[SCI]{Michael Penwarden}
\ead{mpenwarden@sci.utah.edu.}
\author[SOC]{Shandian Zhe}
\ead{zhe@cs.utah.edu}
\author[MATH]{Akil Narayan}
\ead{akil@sci.utah.edu}
\author[SCI]{Robert M. Kirby}
\ead{kirby@cs.utah.edu}

\address[SCI]{School of Computing and Scientific Computing and Imaging Institute, University of Utah, Salt Lake City, UT}
\address[SOC]{School of Computing, University of Utah, Salt Lake City, UT}
\address[MATH]{Department of Mathematics and Scientific Computing and Imaging Institute, University of Utah, Salt Lake City, UT}

\begin{abstract}
Physics-informed neural networks (PINNs) as a means of discretizing partial differential equations (PDEs) are garnering much attention in the Computational Science and Engineering (CS\&E) world.  At least
two challenges exist for PINNs at present:  an understanding of accuracy and convergence characteristics with respect to tunable parameters and identification of optimization strategies that make PINNs as efficient as other computational science tools. The cost of PINNs training remains a major challenge of Physics-informed Machine Learning (PiML) -- and, in fact, machine learning (ML) in general.  This paper is meant to move towards addressing the latter through the study of PINNs on new tasks, for which parameterized PDEs provides a good testbed application as tasks can be easily defined in this context. Following the ML world, we introduce {\em metalearning} of PINNs with application to parameterized PDEs.  By introducing metalearning and transfer learning concepts, we can greatly accelerate the PINNs optimization process.  We present a survey of {\em model-agnostic} metalearning, and then discuss our {\em model-aware} metalearning applied to PINNs as well as implementation considerations and algorithmic complexity. We then test our approach on various canonical forward parameterized PDEs that have been presented in the emerging PINNs literature.

\end{abstract}


\end{frontmatter}


%
\section{Introduction}
\label{sec:introduction}

Physics-informed Machine Learning (PiML) represents the modern confluence of two powerful computational modeling paradigms:  data-intensive machine learning concepts and model-informed
simulation science. Physics-informed Neural Networks (PINNs) \cite{raissi2019physics,raissi2017physicsI,raissi2017physicsII}
as a means of discretizing partial differential equations (PDEs) are garnering much attention in the Computational Science and Engineering (CS\&E) world. PINNs represent a new ``meshfree" discretization methodology built upon deep neural networks (DNNs) and capitalizing on machine learning technologies such as automatic backward differentiation and stochastic optimization \cite{DeepLearning}.  The marriage of computational modeling and machine learning is predicted to transform the way we do science, engineering, and clinical practice \cite{Alber2019}. Some argue that they augment existing items in our computational science toolbox, while others claim they may supplant traditional methods such as finite elements, finite volumes, and finite differences.  Regardless, at least two challenges exist for PINNs: an understanding in a tunable way of their accuracy and convergence characteristics and optimizations that make PINNs as efficient as other computational science tools.  

We focus on the latter issue:  in their current state, PINNs often are far more expensive to train than the computational (often linear algebra-based) solver kernels of other traditional methods.  PINNs proponents
argue that traditional methods, particularly the area of computational linear algebra, have had over forty years of concerted effort expended towards their development and optimization and that the comparison will only be fair when the same amount of energy is used expended towards accelerating PiML. Correspondingly, they argue, it is not surprising that in the nascent field of PiML, the cost of PINNs training remains a major challenge of physics-informed machine learning (PiML) -- and in fact, machine learning (ML) in general -- during these early days of development.

This paper is meant to move towards addressing the optimization challenge previously described through the study of PINNs applied to parameterized PDE. This has been previously addressed in the context metalearning PINN loss functions on parametric PDEs \cite{PSAROS2022111121} whereas we metalearn PINN weight initialization. Recently, \cite{liu2022novel} attempts to metalearn PINN weights with a new Reptile initialization, a gradient-based method similar to MAML. However, in contrast to our approach, they do not perform the more accurate L-BFGS optimization step and results are shown to be unconverged due to the limited number of iterations performed at test time. These results do show improved performance to the start of training, which we also observe for the similar MAML approach. However, we find MAML struggles in the fine-tuned L-BFGS optimizer regime whereas our metalearning method performs better and with reduced cost on the type of training scheme used in this paper. 

Literature exists for solving parametric PDEs in the PiML context \cite{pi-deeponet, long-time_pi-deeponet, deep-op-networks, SUN2020112732, ZHU201956, GAO2021110079, GAO2022114502}, many of which focus on function-to-function maps, in contrast to our approach. They do not attempt to address a broader metalearning problem as well as PINNs directly learning the solution for each task. We emphasize our method is not a replacement to these papers but a complement to existing literature. PINNs also have the benefit of a vast collection of extensions. Beyond the initial collocation version of PINNs, Karniadakis and collaborators have extended these methods to conservative PINNs (cPINNs)  \cite{JAGTAP2020113028}, variational PINNs (vPINNS) \cite{kharazmi2019variational}, parareal PINNs (pPINNs) \cite{MENG2020113250}, stochastic PINNs (sPINNs) \cite{ZHANG2019108850}, fractional PINNs (fPINNs) \cite{pang2018fpinns}, LesPINNs (LES PINNs) \cite{PhysRevFluids.4.034602}, non-local PINNs (nPINNs) \cite{pang2020npinns} and eXtended PINNs (xPINNs) \cite{JagtapK}. In this work, we will focus on application of the original collocation PINNs approach; however, the work presented herein can be applied
to many if not all of these variants.

Following the ML world, we introduce {\em metalearning} of PINNs with application to parameterized PDEs.  By introducing metalearning and transfer learning concepts, we can greatly accelerate the PINNs optimization process.  We present a survey of {\em model-agnostic} metalearning and transfer learning concepts and then discuss our {\em model-aware} specialization of metalearning applied to PINNs.  We provide theoretically motivated and empirically backed assumptions that make our metalearning approach possible. We then test our approach on various canonical forward parameterized PDEs that have been presented in the emerging PINNs literature.

The paper is organized as follows:  In Section \ref{sec:metalearning}, we provide an overview of metalearning and transfer learning and then explain how finding solutions over parameterized PDEs can be viewed as a metalearning problem. In Section \ref{sec:pinns}, we first review the original PINNs collocation approach and provide a summary of current and ongoing PINNs efforts within the field.  Although we focus on the application of our metalearning approach to the collocation version of PINNs, nothing precludes the extension of our work to other PINNs variants with appropriate generalization and minor modifications. In Section \ref{sec:results}, we present our metalearning PINNs approach applied to forward parameterized PDE problems that have been presented in the emerging PINNs literature. Furthermore, we discuss our approach's limitations and assumptions with open methodological challenges to the PINNs and machine learning communities. We conclude in Section \ref{sec:summary} with a summary of our work and a discussion of current challenges and potential future avenues of inquiry and expansion of the concepts presented in this work. Additionally, we provide a survey of two common categories of methods used -- statistical methods for regression and numerical methods for building approximations, which can be viewed in the PINNs context as {\em model-aware} metalearning algorithms in \ref{sec:methods}. We also provide supplementary discussion on theoretical consideratiosn and additional computation results in \ref{sec:theoretical} and \ref{sec:additional} respectively.

\section{Metalearning and Parameterized PDEs}
\label{sec:metalearning}

\renewcommand{\X}{\bXi}	%

\renewcommand{\x}{\btx}
\renewcommand{\y}{\bty}
\renewcommand{\z}{\btz}

\renewcommand{\b}{\bpsi}
\renewcommand{\B}{\bPsi}

\renewcommand{\v}{\bupsilon}

In this section, we first present an overview, from the machine learning perspective, of the emerging areas
of metalearning and transfer learning.  Subsequently in this paper we use the term {\em metalearning} to refer
to both (although we acknowledge that 
there are nuanced and important differences).  We
then discuss how computing solutions of parameterized PDEs can be viewed as a metalearning problem.

\subsection{Overview of Metalearning and Transfer Learning}
\label{sec:overviewmeta}


Metalearning~\cite{schmidhuber1987evolutionary, thrun2012learning} is a class of machine learning methodologies that intends to quickly adapt a learning model to new tasks or environments, mimicking human learning. To this end, metalearning usually extracts some common, important meta information, such as the initial values of the model parameters and other hyperparameters from a family of training tasks that are correlated to new, unseen tasks. This meta information is used to conduct the training on the new tasks, which is expected to accelerate the training and/or improve the performance. Metalearning is a cross-disciplinary research area between multitask learning~\cite{caruana1997multitask} and transfer learning~\cite{pan2009survey,torrey2010transfer}. Current metalearning approaches can be (roughly) classified into three categories: (1) metric-learning methods that learn a metric space (in the outer level), where the tasks (in the inner level) make predictions by simply matching the training points, \eg nonparametric nearest neighbors \cite{koch2015siamese, vinyals2016matching, snell2017prototypical, oreshkin2018tadam, allen2019infinite}; (2) black-box methods that train feed-forward or recurrent neural networks (RNNs) to take the hyperparameters and task dataset as the input and  predict the optimal model parameters or parameter updating rules \cite{hochreiter2001learning, andrychowicz2016learning, li2016learning, ravi2016optimization, santoro2016meta, duan2016rl, wang2016learning, munkhdalai2017meta, mishra2017simple}; and   (3) optimization-based methods that conduct a bi-level optimization, where the inner level is to estimate the model parameters given the hyperparameters and the outer level is to optimize the hyperparameters via a meta-loss~\citep{finn2017model, finn2018learning, bertinetto2018meta, zintgraf2019fast, li2017meta, finn2018probabilistic, zhou2018deep, harrison2018meta}. Other (hybrid) approaches include \cite{rusu2018meta, triantafillou2019meta}, \etc An excellent survey about metalearning for neural networks is given in~\cite{hospedales2020meta}.

\subsection{Model-Agnostic Metalearning}
\label{sec:maml}
We discuss in this section a popular and effective metalearning strategy. 
Suppose we are interested in a family of correlated learning tasks $\Acal$, over which we will employ a machine learning model $\Mcal$, \eg a deep neural network. Note that $\Acal$ can be an infinite set.  A particularly successful metalearning algorithm is model-agnostic metalearning (\maml)~\citep{finn2017model} that aims to find an initialization  $\btheta$ for training $\Mcal$ that can well adapt to all the tasks in $\Acal$. To this end, \maml  samples $N$ tasks $\{\Tcal_{1}, \ldots, \Tcal_{N}\}$ from $\Acal$, and for each $\Tcal_{n}$, collects a dataset $\Dcal_n$. Each  $\Dcal_n$ is further partitioned into a metatraining dataset $\Dcal_n^{\text{tr}}$ and a meta-validation dataset $\Dcal_n^{\text{val}}$. To evaluate the task learning performance with $\btheta$ as the initialization, MAML performs one-step (or a few steps of) gradient descent with the loss on $\Dcal_n^{\text{tr}}$, and then evaluates the loss on $\Dcal_n^{\text{val}}$,
\begin{linenomath}\begin{align}
	\Lcal(\btheta - \eta \nabla \Lcal(\btheta, \Dcal_n^{\text{tr}}), \Dcal_n^{\text{val}}), \label{eq:meta-loss}
\end{align}\end{linenomath}
where $\eta$ is the step size.  This essentially assumes that the model performance after one-step (or a few steps of) update from the initialization can well describe the performance after more thorough training.  Accordingly, MAML minimizes the following meta-objective to optimize $\btheta$, 
\begin{linenomath}\begin{align}
	\btheta^* =  \argmin_{\btheta} \sum_{\Tcal_n \in \Scal} \Lcal\big(\btheta - \eta \nabla \Lcal(\btheta, \Dcal_n^{\text{tr}}), \Dcal_n^{\text{val}}\big). \label{eq:maml}
\end{align}\end{linenomath}

We use MAML as the baseline metalearning method against which to compare in the results section. 

\subsection{Parameterized PDEs as a Metalearning Problem}
\label{sec:parameterPDE}

We now describe how solutions to parameterized PDEs can be viewed as a metalearning problem. In brief, we view different parameter values or regions as a collection of tasks; data for learning parametric behavior is gathered as snapshot (fixed-parameter) solutions to the PDE. Consider a general parameterized nonlinear system of steady-state or transient PDEs of arbitrary order for the dependent scalar variable $u(\x,t;\bxi)$,
\begin{linenomath}\begin{equation}\label{eq:pde}
\begin{cases}
\frac{\partial}{\partial t} u(\x,t;\bxi)  + \mathcal{F}\left( u(\x,t;\bxi)\right) = \mathcal{S}(\x,t;\bxi),  & \Omega \times [0, T] \times \mathcal{X}, \\
\mathcal{B}(u; \bxi) = 0, & \partial \Omega \times [0, T] \times \mathcal{X}, \\
u(x,0;\bxi) = u_0(x;\bxi), & \Omega \times \{t=0\} \times \mathcal{X}, \\
\end{cases}
\end{equation}\end{linenomath}
where $\mathcal{F}$ is a nonlinear operator that may contain parameters $\bxi \in \mathcal{X} \subset \mathbb{R}^m$.  
$\mathcal{S}$ is the source term/function, $\Omega$ and $T$ are the spatial and temporal domain of interest, $\mathcal{B}$ is the boundary condition operator also potentially parameterized via $\bxi$, and $u_0(\x,\bxi)$ parameterizes the initial condition.
The variable $\x \in \Omega \subset \mathbb{R}^{s}$ is the spatial coordinate of an $s$-dimensional space and $t \in [0, T]$ represents the temporal variable.
The PDEs can be fully nonlinear and parameterized in an arbitrary fashion (including the initial and boundary conditions); we assume the parametric solution map is well-posed, \ie the solution map $\bxi \mapsto u(\cdot,\cdot;\bxi)$ is a well-defined map from $\mathcal{X}$ to an appropriate function space over $\Omega \times [0,T]$.  Conceptually, based upon the reasoning in \cite{EIM2004}, we seek to define a manifold of solutions over a parameter space of dimension $m$.  


For a given parameter $\bxi$, the system \eqref{eq:pde} can be numerically solved for $u(\x, t)$ typically using an approximate solver, e.g., based upon finite difference, finite element or finite volume methods.  For applications such as uncertainty quantification (UQ) (where $\bxi$ encodes the randomness) and design optimization (where $\bxi$ encodes the design parameters), a large number of such forward evaluations is required. Direct implementation via a numerical solver can be computationally prohibitive, especially in the many-query contexts of design and UQ.  In the subsequent section (Section \ref{sec:pinns}), we will highlight the use of PINNs as an alternative numerical solver for these purposes, which will be our computational strategy for sampling from the solution manifold.

A direct approximation of $u(\x,t;\bxi)$ is difficult due to the number of samples we need to cover the response surface for such a high-dimensional input space problem~\citep{higdon2008computer}. Following the pioneer work of Higdon et al.~\cite{higdon2008computer}, we can record values at specified (fixed) spatial locations $ \x_1, \dots, \x_{N_x} $ and temporal locations $t_1,\dots, t_{N_t} $ to provide a discrete approximation of $u(\x, t;\bxi)$. 
With the recording coordinates,  our quantity of interest becomes a vectorial function of the PDE parameters
\begin{linenomath}$$\y(\bxi) = \left(u(\x_1,t_1;\bxi),\dots,u(\x_{N_x},t_1;\bxi),u(\x_1,t_2;\bxi),\dots,u(\x_{N_x}, t_{N_t}; \bxi) \right)^\top \in \mathbb{R}^d,$$\end{linenomath}
where $\x_i$ is a spatial coordinate of a regular/irregular grid, and $d=N_x\times N_t$ is the total number of spatial-temporal grid points.\footnote{We assume that $u(x, t; \bxi)$ is scalar-valued, but straightforward extensions can be considered for systems of PDEs.}
Note that although we use the language of solution evaluation, all of these definitions can be modified to denote spatial and temporal degrees of freedom
more broadly (e.g. Fourier modes, etc.).  In the subsequent section (Section \ref{sec:pinns}), these evaluation points will be related to the collocation points of the PINNs 
solution.

Following \cite{EIM2004}, the challenge in general 
becomes how to approximate the mapping $\y(\cdot):\mathcal{X} \to \RR^{d}$ with a limited computational budget.  There is a rich 
literature around the topic of approximating the solution manifold over the parameters; we refer the reader to \cite{ReducedBasis}
which both surveys and summarizes these methods.  Furthermore, in recent years, methods from both the statistical and/or machine
learning communities have been proposed, e.g. ~\cite{perdikaris2017nonlinear,cutajar2019deep,WXiu2021}.

Using the language of Section \ref{sec:overviewmeta}, we now recast the problem statement above into a {\em metalearning} problem.
Assume we start with a sampling of the parameter space ${\bxi}_i \in \mathcal{X}, \, i=1,\dots, K$ at which we evaluate $u(\x,t;\bxi)$ (or its discretized version $\y(\bxi)$).  
The particular choice of sampling will depend on what mathematical or statistical properties may be needed in subsequent computation.
Since different parameters denote different specific PDEs, each parameter choice, in the extreme, can be considered a {\em task}.  
However, often non-overlapping subsets of the parameter space $\mathcal{X}_i \subset \mathcal{X}, \, i = 1,\ldots,K$, the union of which amounts to the full parameter space, represent a finite collection of tasks.  In this case, tasks denote regions of the parameter space
over which the mathematical characteristics of the PDE remain the same.  For example, consider the compressible Navier-Stokes equations 
for which Mach number $Ma$ and Reynolds number $Re$ denote the parameter space.  Partitioning values of $Ma$ according to the boundary $Ma=1$ may clearly 
denote two different parameter spaces for which the mathematical properties of the PDE differ.  Based upon these observations, one
can appreciate that although significant research within the machine learning domain has been placed on {\em model-agnostic} 
metalearning, using this paradigm we can employ various reduced-order modeling techniques to generate {\em model-aware}
metalearning. In the next section, we will first review summarize PINNs and how it fits into this framework, and then state 
both the theoretical and implementation considerations that are necessary to successfully employ this view of parameterized PDE
metalearning.

\noindent \textbf{Remark.} We emphasize that our approach is {\em model-aware} in the sense that we leverage prior knowledge of the task class, in this case parametric domains in which the mathematical characteristics of the PDE remain the same, to help metalearn PINN weight initalizations.

\section{Physics-Informed Neural Networks (PINNs)}
\label{sec:pinns}

In this section, we first present a review of physics-informed neural networks (PINNs), with an emphasis on the original collocation PINNs approach which we use in this work. We also provide a brief summary of current and ongoing PINNs efforts within the field -- many if not all of which might benefit from the metalearning approach presented herein. We then present the application of PINNs metalearning to parameterized PDEs. Finally, we provide a summary of the implementation considerations that are required as well as the algorithmic complexity.

\subsection{Review of Physics-Informed Neural Networks}\label{ssec:pinns}

 Physics-Informed Neural Networks (PINNs) were originally proposed by Karniadakis and co-workers \cite{raissi2019physics,raissi2017physicsI,raissi2017physicsII} 
 as a neural network based alternative to traditional PDE discretizations. In the original PINNs work, when presented with a PDE specified 
 over a domain $\Omega$ with boundary conditions on $\partial \Omega$ and initial conditions at $t=0$ (in the case of time-dependent PDEs),
 the solution is computed (i.e. the differential operator is satisfied) 
 as in other mesh-free methods like RBF-FD \cite{ShankarWFK1,ShankarWFK2} at a collection of collocation points.  
 First, we re-write our PDE system in residual form as 
 $R(u) = \mathcal{S} - \frac{\partial}{\partial t} u - \mathcal{F}(u)$, where $\mathcal{S}$ and $\mathcal{F}$ are as defined in \eqref{eq:pde}.
 The PINNs formulation is expressed as follows:  Given a neural network function $\tilde{u}(\x,t;\underbar{w})$ with specified activation functions
and a weight matrix $\underbar{w}$ denoting the degrees of freedom derived from the width and depth of the network,  find 
$\underbar{w}$ that minimizes the loss function:

\begin{equation}
MSE = MSE_u + MSE_R
\end{equation}

\noindent where

\begin{eqnarray}
MSE_u &=& \frac{1}{N_u} \sum_{i=1}^{N_u} \| \tilde{u}(x_u^i,t_u^i;\underbar{w}) - u^i \|^2  \\
MSE_R &=& \frac{1}{N_R} \sum_{i=1}^{N_R} \| R(\tilde{u}(x_R^i,t_R^i) \|^2 \,\,\, 
\end{eqnarray}

\noindent where $\{x_u^i,t_u^i,u^i\}_{i=1}^{N_u}$ denote the initial and boundary training data on $u(\x,t)$ 
and $\{x^i_R,t^i_R\}_{i=1}^{N_R}$ specify the collocation points for evaluation of the collocating residual term $R(\tilde{u})$.  
The loss $MSE_u$ corresponds to the initial and boundary data while $MSE_R$ enforces the structure imposed by the
differential operator at a finite set of collocation points.  This loss-function modified minimization approach fits naturally 
into the traditional deep learning framework \cite{DeepLearning}.  Various optimization choices are available including
stochastic gradient descent (SGD), L-BFGS, etc.  The result of applying this procedure is a neural network $\tilde{u}(\x,t;\underbar{w})$
that attempts to minimize through a balancing act the strong imposition of the initial and boundary conditions and satisfaction
of the PDE residual.  Note that this statement does not immediately connect to the approximation error $\| u(\x,t) - \tilde{u}(\x,t;\underbar{w})\|$;
however, consistency and convergence are items of current research (e.g. \cite{Shin20}).



\subsection{Application of PINNs Metalearning to Parameterized PDEs}
\label{ssec:pinn-meta}

This section provides a connection between metalearning and the application of parameterized PDEs that we will exploit for PINNs algorithms. To begin, we provide a more detailed overview of the anatomy of a PINN that complements the high-level discussion of Section \ref{ssec:pinns}.

In the context of parameterized PDEs (Section \ref{sec:parameterPDE}), we consider an arbitrary but fixed value $\bxi$ of the parameter. A PINN emulator is the map $(\x,t) \mapsto \tilde{u}(\x, t; \underbar{w})$, where we assume that the weights $\underbar{w}$ are trained as discussed in Section \ref{ssec:pinns}. The function $\tilde{u}$ is a neural network, i.e., an iterative compositions of affine maps and an activation function $\sigma$; in the special case of PINNs this can be summarized as
\begin{linenomath}\begin{align*}
h_0 &= (\x,t) &\\
h_{j} &= \sigma\left( W_{i} h_{j-1} + b_{j}\right) & j = 1, \ldots, \ell+1 \\
\tilde{u} &= h_{\ell+1} 
\end{align*}\end{linenomath}
where $h_j \in \mathbb{R}^{n_j}$ are hidden states, $\sigma: \mathbb{R} \rightarrow \mathbb{R}$ is an activation function that is applied componentwise to vectors, $b_j \in \mathbb{R}^{n_j}$ are bias vectors, and $W_j \in \mathbb{R}^{n_j \times n_{j-1}}$ are weight matrices. The network has $\ell$ hidden layers, and the width of layer $j$ is $n_j$.
The input layer $(\x,t)$ is of width $n_0 = p+1$, and the output layer is of width $n_{\ell+1} = 1$. With $\underbar{W}_j \coloneqq \left[ W_j\;\; b_j \right] \in \mathbb{R}^{n_j \times (n_{j-1}+1)}$ the concatenation of weights and biases in layer $j$, then we collect all the parameters of the neural network in the \textit{weight} vector,
\begin{linenomath}\begin{align*}
  \underbar{w} &\coloneqq \left( \mathrm{vec}(\underbar{W}_1)^\top, \;\; \ldots, \;\; \mathrm{vec}(\underbar{W}_{\ell+1})^\top \right)^\top \in \mathbb{R}^{M_\ell}, & M_\ell &= \sum_{j=1}^{\ell+1} n_j (n_{j-1}+1).
\end{align*}\end{linenomath}

We have assumed a simple network architecture above, e.g., fully connected and with the same activation function for all neurons. However, the discussion can be appropriately extended to more customized network architectures. In the future, we will refer to $\underbar{w}$ as the weights that are trained through the procedure in Section \ref{ssec:pinns}. 

In the context of metalearning, the map $\bxi \mapsto \underbar{w}$ is the task associated to parameter $\bxi$. With this in mind, we will write $\underbar{w} = \underbar{w}(\bxi)$ to explicitly communicate that the trained weights depend on $\bxi$. The metalearning problem here is, given a parameter $\bxi$, identify a set of weights that is well-adapted to the task of identifying $\underbar{w}(\bxi)$. More precisely, given $K$ sample task parameters $\{\bxi_1, \ldots, \bxi_K\} \subset \mathbb{R}^m$, compute a task predictor $\hat{\underbar{w}}(\bxi)$ satisfying $\hat{\underbar{w}}(\bxi) \approx \underbar{w}(\bxi)$ for all parameters $\bxi \in \mathcal{X}$. In this paper, we build $\hat{\underbar{w}}$ as a linear map,
\begin{linenomath}\begin{align}\label{eq:linear-approx}
  \hat{\underbar{w}}(\bxi) = \sum_{k=1}^K \b_k c_k(\bxi),
\end{align}\end{linenomath}
where $\b_k$ are vectors in $\mathbb{R}^{M_\ell}$, and $c_k(\cdot)$ are coefficient functions. We will utilize various approaches that will specify the vectors $\b_k$ and coefficients $c_k$ via available sample task outputs $\{\underbar{w}(\bxi_1), \ldots, \underbar{w}(\bxi_N)\}$. Our mathematical approaches will treat $c_k$ as deterministic functions, whereas statistical approaches treat $c_k$ as random variables. 

\subsubsection{Implementation Considerations}

The PINN metalearning approach laid out in Section \ref{ssec:pinn-meta} contains three steps, summarized as: (1) sample the parameter space appropriately for the determined weight prediction method; (2) evaluate the PINN at the sampled locations and store trained weights; (3) construct the weight prediction model. When queried at a new location, this predictive method should output a reasonable estimation of what the trained PINN weights would be, thus reducing the optimization time. Based on this, we must decide how to sample the space and what prediction method to use. In this paper, we present five methods with appropriate sampling for each, implying that we are not advocating any particular method over another but rather to show there are many ways this can be approached. We also compare this to the standard randomization of weights, MAML, and a zeroth-order model in which we initialize with the trained weights at the center of the parameter space $\mathcal{X}$. Any method or sampling could be chosen to replace these steps.
 
Additionally, it is essential to ensure the weights vary smoothly across $\mathcal{X}$ as the methods we are using assume smoothness. We have found that initializing the training sets with the trained weights at the center of the parameter space appears to empirically ensure smoothness for one type of PDE task family as mentioned in Section \ref{sec:parameterPDE}. We are attempting to start the weights optimization in the same basin of attraction for all samples by doing this. We assume that by doing this, we obtain smooth weights over $\mathcal{X}$, but this is not guaranteed. Alternatively, initializing with random weights ensures one will fall into many different regions of attraction, and the weights will likely not vary smoothly, and so metalearning with our strategy could suffer in accuracy. Future work will be to develop a more sophisticated way to ensure this condition as well as to discover task boundaries so that the method can be task-agnostic. 


\subsubsection{Algorithmic Complexity}

To aid in evaluating the algorithmic complexity and reproducibility of our approach, we provide Algorithm \ref{meta-algo}. The primary consideration when evaluating the cost of our approach is the number of training points $K$ as we must train PINNs to obtain the weight data for each point. In this paper, we set K across for all methods as the size of the hyperbolic cross set, which is dependent on the parameter dimension $m$ of $\mathcal{X} \subset \mathbb{R}^m$ and the order of the hyperbolic cross set $d$. These points are used in conjunction with the polynomial least-squares method, and for the other methods, we use a Latin hypercube sampling of size $K$. We use the same $K$ value for both sampling methods to make a fair comparison, but one can choose any $K$. Finding a balance of cost associated with solving $K$ PINNs to construct the models with and the cost savings of the models during testing is the main decision complexity in our approach.

\begin{algorithm}
\caption{Metalearning PINN Weights Method} 
\label{alg1}
\begin{algorithmic}
    \STATE Given appropriate PDE set-up information (i.e. spatial domain $\Omega$, temporal domain $T$ , parameter space $\mathcal{X}$,  etc.)
    \STATE Define PINN hyper-parameters (size and optimization) for $\tilde{u}$
    \STATE Set $\bxi$ as the appropriate sampling of $\mathcal{X}$ with size $K$ for the weight prediction model used
    \STATE Let $\bxi^{\cal{C}}$ be the centroid of $\mathcal{X}$
    \STATE Store trained weights $\underbar{w}(\bxi^{\cal{C}})$ from $\tilde{u}(\x,t;\bxi^{\cal{C}})$
    \STATE Assume trained PINN weights $\underbar{w}(\bxi)$ are smooth across $\mathcal{X}$ if initialized with $\underbar{w}(\bxi^{\cal{C}})$ 
    \FOR {j = 1 \TO $K$}
    	\STATE Initialize $\tilde{u} \: $weights$ \: \underbar{w}(\bxi_j)$ with $\underbar{w}(\bxi^{\cal{C}})$
    	\STATE Store vectorized trained weights $\underbar{w}(\bxi_j)$ from $\tilde{u}(\x,t;\bxi_j)$
    \ENDFOR
    \STATE Construct weight prediction models $\hat{\underbar{w}}(\bxi)$ (GPs, RBF, etc.) from stored vectorized weight data $\underbar{w}(\bxi)$
    \STATE Predict PINN weights for any point $\hat{\underbar{w}}(\bxi^*)$ to initialize a PINN with, then train
\end{algorithmic}
\label{meta-algo}
\end{algorithm}


%
\section{Results and Discussion}
\label{sec:results}

In this section, we demonstrate the efficacy of our metalearning approach on various representative forward PDE problems. The examples below are extensions of the test problems proposed in \cite{raissi2019physics,Yang2020,PSAROS2022111121}. There are two types of spatiotemporal domains we consider: (i) the Burgers' and (nonlinear) Heat equations in which we have a spatial and temporal dimension, and the (ii) Allen-Cahn, Diffusion-Reaction, and linear parametric heat equations in which we have two spatial dimensions. Regarding the parametric dimension, we provide results for 1D, 2D, 6D, and 10D cases. We also emphasize these are \textit{intrinsic} not \textit{extrinsic} dimensions and should not be compared with methods that deal in $\mathcal{O}(100-1000)$ \textit{extrinsic} dimensions as those would in practice be reduced to a much smaller \textit{intrinsic} dimensional problem.

Lastly, a comprehensive review of methods for surrogate modeling are provided in \ref{sec:methods} which we use to estimate the weight initializations given observations across the parametric domain.

\subsection{Experimental settings}

For PDE type (i), the PINN solution test points are a uniform grid of size $256 \times 100$ in space and time respectively. For type (ii), we use a uniform grid of size $128 \times 128$. These test points are where we compare between the exact solution and the PINN solution. To train the PINN we use $100$ uniformly sampled boundary/initial value points for the $MSE_u$ portion of the loss, and $10,000$ collocation points using Latin hypercube sampling (LHS) for the $MSE_R$ portion as described in Section \ref{ssec:pinns}. In all PINN architectures, we employ fully-connected feed-forward neural networks with tanh activation functions.  The width and depth parameters will be specified per experiment below. We emphasize that the choice of width and depth are sufficient to solve the PDEs to a reasonable accuracy. PINNs and most physics-informed methods suffer from an optimization error ceiling: A larger network does not in practice translate directly to lower accuracy due to optimization. For Burgers, we provide a comparison of network sizes and show our method does not suffer from needing to initialize more learnable parameters and have chosen network sizes which show a accuracy discrepancy for this purpose \cite{PENWARDEN2022110844}. However, to go beyond this to very large network is unnecessary as one would hit an optimization ceiling at around $10^{-3}$ to  $10^{-4}$ relative $L^2$ error. Lastly, the values in each table are presented as the empirical mean and standard deviation over the test set, which is a sampling of problems over the defined parametric domain. These experiments were run on an Intel Core i7-5930K processor with the Windows 10 OS.


The table columns are relative $L^2$ error after 500 iterations of Adam optimization, relative $L^2$ error after L-BFGS optimization, and the time taken to optimize L-BFGS for a certain tolerance condition. 
The relative $L^2$ error over the test points is given by:
\begin{linenomath}\begin{align}
	\frac{||u-\tilde{u}||_2}{||u||_2}
\end{align}\end{linenomath}
The sequential optimization with Adam followed by L-BFGS is standard approach in PINNs. \cite{Shin20}. The L-BFGS conditions are $10^{-6}$ termination tolerance on first order optimality and $10^{-9}$ termination tolerance on weight vector changes with a learning rate of 0.1.

For MAML, we train with 1,000 epochs and Adam with learning rate $10^{-3}$, which takes approximately 1,000 seconds wall-clock-time to run for the Burgers problem The inner optimization uses one-step gradient decent and step-size $10^{-3}$. The data used has the same number of boundary points, collocation points, and tasks for the training and validations sets as the other predictive methods presented.

Lastly, we point out that occasionally we train ''bad" PINNs where they do not converge relative to their loss function. This has been observed only for random and RBF Gaussian kernel initialization. The random initializations are a byproduct of standard PINNs, and the unconverged runs can be identified with a large loss and re-run. For our predictive RBF Gaussian kernel method, re-running does not help as it provides the same initialization for each point so these were identified via loss and removed. We further discuss RBF Gaussian kernel in the contexts of the results as being a poor method for our metalearning approach. Since the GP approach is a generalization of the RBF approach (it optimizes hyperparameters), then our results suggest the predictable result that optimizing these hyperparameters, which is common in GP methods, appears better than empirically choosing them (e.g., based on fill distances), which is a common workflow in RBF methods.

\subsection{1D Burgers Equation}
\label{ssec:burgers}

We consider the following 1D viscous Burgers equation:

\begin{linenomath}\begin{equation}
\frac{\partial u}{\partial t} + \frac{1}{2}\frac{\partial}{\partial x} \left( u^2 \right) = \nu \frac{\partial^2 u}{\partial x^2}, \, x \in \Omega, \, t \in [0,T]
\end{equation}\end{linenomath}

\noindent where $\Omega \times [0,T] =  [-1,1], \times [0,1]$ with the viscosity $\nu \in [0.005,0.05]$ and initial condition $u(x,0) = -sin(x\pi)$. For evaluation of the error, we compute the exact solution derived using Cole's transformation computed with Hermite 
integration \cite{Patera86}. We initially randomly sample our parameter space with $16$ PINNs approximations containing $5$ hidden layers of width $5$. The size $16$ derives from the amount of points with the degree 5 hyperbolic cross as described in \ref{sec:cross-sets}.

\begin{center}
\captionof{table}{We sample the PDE parameter space uniformly with 32 points to compute the values in the table.} 
\scalebox{0.75}{
\begin{tabular}{| c | c | c | c |}
\hline
Initialization Methods & Error after Adam ($10^{-3}$) & Error after L-BFGS ($10^{-3}$) & L-BFGS Time (s) \\
\hline
Random & $618.1 \pm 177.7$ & $5.1 \pm 6.0$ & $146 \pm 68$ \\

MAML &  $107.7 \pm 40.5$ & $7.4 \pm 5.7$ & $184 \pm 66$ \\

Center & $58.6 \pm 103.7$ & $6.4 \pm 9.1$ & $63 \pm 51$ \\

Multitask & $58.5 \pm 84.1$ & $5.2 \pm 8.5$ & $53 \pm 34$ \\

LMC & $292.6 \pm 98.7$ & $5.0 \pm 6.7$ & $58 \pm 36$ \\

Spline & $40.9 \pm 65.5$ & $5.9 \pm 9.1$ & $40 \pm 28$ \\

RBF (cubic) & $26.1 \pm 57.6$ & $5.4 \pm 9.3$ & $27 \pm 40$ \\

RBF (gaussian) & $54.2 \pm 93.4$ & $5.1 \pm 8.2$ & $39 \pm 25$ \\

RBF (multiquadric) & $28.9 \pm 63.3$ & $4.6 \pm 6.6$ & $26 \pm 29$ \\

Polynomial & $23.5 \pm 41.4$ & $5.1 \pm 10.1$ & $39 \pm 37$ \\
\hline
\end{tabular}}
\label{table:burgers_performance}
\end{center}

In Table \ref{table:burgers_performance} we first observe that MAML improves the error after Adam optimization compared to randomized weights; however it worsens the convergence with L-BFGS. Note that Adam and SGD are first-order methods whereas L-BFGS is a second-order method. In PINNs, we further optimize with L-BFGS after Adam which evidently is an issue when using MAML. This may be due to MAML using first-order gradient descent in the inner training loop (see Equation \ref{eq:meta-loss}), which is inconsistent with switching to second-order methods in practice. MAML appears to do worse than randomized weights in the context of this sequential training in regard to L-BFGS time. In the context of training without L-BFGS, we note that our metalearning method still outperforms MAML after only performing the Adam iterations. We also note that MAML has greater offline cost than our simple Center initialization method which only requires one PINN run and a comparable cost to the other methods for the given settings. In our interpolation methods, we can also see that they greatly improve the time it takes to reach the L-BFGS tolerance, by around a factor of five, while achieving the same accuracy. One trend we note throughout all problems presented is how well initializing with a run at the center of the parametric domain does for its minimal cost, this will be further discussed. 

\textbf{Remark:} There will be some performance variance in these methods depending on the randomization of the weights for the center run. This run is used to initialize the data for interpolation, so it follows that it affects the final outcome. We recommend running the parametric center PINN a few times and taking the most accurate of the learned weights to address this issue. This is also an issue for MAML as it also is dependent on the random weights at which it starts before meta training.

Next, we use a larger PINN containing $8$ hidden layers of width $10$ to see how this affects the methods. 

\begin{center}
\captionof{table}{We sample the PDE parameter space uniformly with 32 points to compute the values in the table.} 
\scalebox{0.65}{
\begin{tabular}{| c | c | c | c | c |}
\hline
Initialization Methods & Error after Initializing ($10^{-3}$) & Error after Adam ($10^{-3}$) & Error after L-BFGS ($10^{-3}$) & L-BFGS Time (s) \\
\hline
Random & $1245.5 \pm 287.3$ & $174.9 \pm 130.4$ & $1.2 \pm 1.8$ & $105 \pm 29$ \\

MAML & $334.6 \pm 37.8$ &  $92.3 \pm 30.3$ & $1.7 \pm 3.2$ & $125 \pm 38$ \\

Center & $70.6 \pm 41.6$ & $14.9 \pm 38.1$ & $0.9 \pm 0.5$ & $56 \pm 48$ \\

Multitask & $96.8 \pm 80.9$ & $2.0 \pm 2.3$ & $0.7 \pm 0.2$ & $24 \pm 21$ \\

LMC & $246.7 \pm 32.4$ & $7.8 \pm 17.5$ & $0.8 \pm 0.4$ & $40 \pm 27$ \\

Spline & $22.7 \pm 8.6$ & $1.4 \pm 1.7$ & $0.7 \pm 0.2$ & $10 \pm 11$ \\

RBF (cubic) & $4.8 \pm 4.4$ & $1.5 \pm 1.9$ & $0.8 \pm 0.4$ & $7 \pm 13$ \\

RBF (gaussian) & $20.0 \pm 19.7$ & $10.6 \pm 18.2$ & $0.8 \pm 0.5$ & $36 \pm 30$ \\

RBF (multiquadric) & $6.1 \pm 7.0$ & $1.2 \pm 1.2$ & $0.8 \pm 0.4$ & $7 \pm 8$ \\

Polynomial & $21.4 \pm 10.6$ & $1.2 \pm 1.9$ & $0.7 \pm 0.3$ & $7 \pm 16$ \\
\hline
\end{tabular}}
\label{table:burgers_performance_large}
\end{center}

In Table \ref{table:burgers_performance_large} we observe that increasing the neural network size not only improves the accuracy but also decreases the time. The analysis of the expressibility of a network is not in the scope of this paper; however, we report this experiment to show our method works regardless of the selection of PINN size. Additionally, we can see that with this setting, most of the methods achieve $10^{-3}$ error with only Adam optimization, meaning one need not employ the more costly refinement of L-BFGS if the initialization is good. In fact, we also include the column for error after initializing before any optimization is performed. Here we can see the best possible case; we have achieved $10^{-3}$ error without any training whatsoever for RBF cubic and multiquadratic kernels meaning the network is initialized very well for the PDE parameters.  

We observe that, among the RBF approaches, the Gaussian kernel is empirically the worst metalearning strategy. Such poor performance could manifest if the scale parameter of the Gaussian kernel is chosen poorly. In our experiments, we use values that are deterministically computed based on fill distances of the parametric grid, which is known empirically to be a reasonable choice \cite{fasshauer_meshfree_2007}. However, the consistently poor performance of Gaussian RBFs in all the remaining examples we have tested suggests that the Gaussian kernel is simply a poor choice of kernel with parameterized PDE metalearning problems.

\subsection{1D nonlinear Heat Equation}
\label{ssec:heat}

We consider the following 1D nonlinear PDE:

\begin{linenomath}\begin{equation}
\frac{\partial u}{\partial t} -\lambda \frac{\partial^2 u}{\partial x^2} + k \,tanh(u) = f, \, x \in \Omega, \, t \in [0,T]
\end{equation}\end{linenomath}

\noindent where $\Omega \times [0,T] = [-1,1] \times [0,1]$ and where $\lambda \in [1,\pi]$ and $k\in[1,\pi]$ are positive constants.  
We specify an exact solution of $u(x,t;\lambda,k) = k \,sin(\pi x)\, exp(-\lambda kx^2)\,exp(-\lambda t^2)$ and derive the corresponding form of the forcing $f$. We initially randomly sample our parameter space with $22$ PINNs approximations 
containing $5$ hidden layers of width $10$.   

\begin{center}
\captionof{table}{We sample the PDE parameter space uniformly with 64 points to compute the values in the table.} 
\scalebox{0.75}{
\begin{tabular}{| c | c | c | c |}
\hline
Initialization Methods & Error after Adam ($10^{-3}$) & Error after L-BFGS ($10^{-3}$) & L-BFGS Time (s) \\
\hline
Random & $465.1 \pm 218.2$ & $5.5 \pm 3.7$ & $178 \pm 37$ \\

MAML &  $384.5 \pm 167.3$ & $6.2 \pm 6.0$ & $211 \pm 58$ \\

Center & $38.4 \pm 28.8$ & $4.6 \pm 2.8$ & $102 \pm 40$ \\

Multitask & $16.7 \pm 14.0$ & $4.6 \pm 3.2$ & $60 \pm 38$ \\

LMC & $20.9 \pm 16.6$ & $4.6 \pm 2.9$ & $66 \pm 43$ \\

Spline & $10.8 \pm 18.2$ & $4.7 \pm 3.5$ & $43 \pm 43$ \\

RBF (cubic) & $28.5 \pm 82.7$ & $4.8 \pm 2.8$ & $42 \pm 38$ \\

RBF (gaussian) & $219.0 \pm 381.1$ & $4.8 \pm 2.8$ & $115 \pm 71$ \\

RBF (multiquadric) & $11.5 \pm 18.7$ & $4.3 \pm 2.6$ & $33 \pm 24$ \\

Polynomial & $7.4 \pm 7.1$ & $4.6 \pm 2.7$ & $30 \pm 24$ \\
\hline
\end{tabular}}
\label{table:heat_performance}
\end{center}

In Table \ref{table:heat_performance} we see considerable improvement in optimization with our initialization procedure. We again observe $10^{-3}$ errors without any need for L-BFGS, in this case with the polynomial interpolation method. 

We now consider the same problem but with a degree 2 hyperbolic cross, equating to $13$ PINN approximations instead of $22$ for degree 5. Because of the conditions for the smooth bivariate spline, a second-order spline is used instead of cubic as described in \ref{ssec:spline}.

\begin{center}
\captionof{table}{We sample the PDE parameter space uniformly with 64 points to compute the values in the table.} 
\scalebox{0.75}{
\begin{tabular}{| c | c | c | c |}
\hline
Initialization Methods & Error after Adam ($10^{-3}$) & Error after L-BFGS ($10^{-3}$) & L-BFGS Time (s) \\
\hline
Random & $529.5 \pm 465.2$ & $5.2 \pm 3.9$ & $156 \pm 42$ \\

MAML & $455.0 \pm 180.0$ & $4.5 \pm 3.1$ & $188 \pm 43$ \\

Center & $38.8 \pm 30.7$ & $4.5 \pm 2.4$ & $96 \pm 41$ \\

Multitask & $18.0 \pm 18.1$ & $4.8 \pm 3.6$ & $49 \pm 30$ \\

LMC & $26.8 \pm 28.3$ & $4.9 \pm 3.3$ & $59 \pm 28$ \\

Spline & $7.7 \pm 6.2$ & $4.7 \pm 2.4$ & $29 \pm 30$ \\

RBF (cubic) & $71.2 \pm 143.2$ & $4.3 \pm 2.0$ & $64 \pm 58$ \\

RBF (gaussian) & $216.5 \pm 320.5$ & $5.4 \pm 4.0$ & $104 \pm 72$ \\

RBF (multiquadric) & $13.9 \pm 19.5$ & $4.4 \pm 2.2$ & $35 \pm 30$ \\

Polynomial & $7.3 \pm 4.7$ & $4.6 \pm 3.2$ & $38 \pm 30$ \\
\hline
\end{tabular}}
\label{table:heat_performance_deg2}
\end{center}

In Table \ref{table:heat_performance_deg2} we observe that lowering the number of data points used for interpolation from 22 to 13 has no significant effect on our method. In fact, the optimization here is slightly better but within reasonable variation for stochastic processes such as training a neural network and variance in sampling locations. We do not claim that fewer points work better but include this experiment to show that our method is not dependent on large amounts of data. We forgo MAML comparisons going forward as this approach empirically struggles with parametric PDEs in the context of PINNs and, specifically, L-BFGS optimization.

\subsection{2D nonlinear Allen-Cahn Equation}

We consider the following 2D nonlinear Allen-Cahn equation, which is a widely used model for multi-phase flows:

\begin{linenomath}\begin{equation}
\lambda \left( \frac{\partial^2 u}{\partial x^2} + \frac{\partial^2 u}{\partial y^2}\right) + u\left(u^2-1\right) = f, \, (x,y)\in\Omega
\end{equation}\end{linenomath}

\noindent where $\Omega = [-1,1]^2$ 
and where $\lambda \in (0,\pi]$ represents the mobility and $u$ denotes the order parameter, which denotes the different phases.  
We specify an exact solution as $u(x,y;\lambda) = exp(-\lambda (x+0.7)) \,sin(\pi x) \,sin(\pi y)$ and derive the corresponding form of the forcing $f$. We initially randomly sample our parameter space with $16$ PINNs approximations containing $8$ hidden layers of width $10$.  

\begin{center}
\captionof{table}{We sample the PDE parameter space uniformly with 32 points to compute the values in the table.} 
\scalebox{0.75}{
\begin{tabular}{| c | c | c | c |}
\hline
Initialization Methods & Error after Adam ($10^{-3}$) & Error after L-BFGS ($10^{-3}$) & L-BFGS Time (s) \\
\hline
Random & $1905.5 \pm 1591.1$ & $14.5 \pm 11.9$ & $469 \pm 161$ \\

Center & $173.3 \pm 172.4$ & $12.3 \pm 4.6$ & $201 \pm 83$ \\

Multitask & $43.3 \pm 45.7$ & $11.5 \pm 4.7$ & $120 \pm 44$ \\

LMC & $57.2 \pm 44.6$ & $11.4 \pm 4.0$ & $120 \pm 21$ \\

Spline & $27.6 \pm 35.8$ & $11.4 \pm 4.0$ & $63 \pm 35$ \\

RBF (cubic) & $21.9 \pm 20.1$ & $11.6 \pm 4.3$ & $64 \pm 42$ \\

RBF (gaussian) & $105.0 \pm 162.1$ & $12.7 \pm 5.2$ & $159 \pm 103$ \\

RBF (multiquadric) & $19.1 \pm 13.5$ & $12.0 \pm 5.3$ & $68 \pm 46$ \\

Polynomial & $30.6 \pm 36.7$ & $11.5 \pm 4.2$ & $44 \pm 19$ \\
\hline
\end{tabular}}
\label{table:allen_performance}
\end{center}

In Table \ref{table:allen_performance} we observe similar trends to the previous two equations. However, we can see the problem is substantially more difficult based on the post L-BFGS accuracy and cost. We again note that by simply initializing with the center run, we gain considerable improvement, regardless of the difficulty of the problem.  

\subsection{2D nonlinear Diffusion-Reaction Equation}

We consider the following 2D nonlinear diffusion-reaction equation:
\begin{linenomath}\begin{equation}
\lambda \left( \frac{\partial^2 u}{\partial x^2} + \frac{\partial^2 u}{\partial y^2}\right) + k \left(u^2\right) = f, \, (x,y)\in\Omega
\label{eqn:diffusion-reaction}
\end{equation}\end{linenomath}

\noindent where $\Omega = [-1,1]^2$.  
Here $\lambda \in [1,\pi]$ represents the diffusion coefficient and $k\in[1,\pi]$ represents the reaction rate and $f$ denotes the source term. We specify an exact solution as $u(x,y;\lambda,k) = k \, sin(\pi x)\, sin(\pi y) \,exp(-\lambda \sqrt{(kx^2+y^2)})$ and derive the corresponding form of the forcing $f$. We initially randomly sample our parameter space with $22$ PINNs approximations containing $8$ hidden layers of width $20$.   

\begin{center}
\captionof{table}{We sample the PDE parameter space uniformly with 64 points to compute the values in the table.} 
\scalebox{0.75}{
\begin{tabular}{| c | c | c | c |}
\hline
Initialization Methods & Error after Adam ($10^{-3}$) & Error after L-BFGS ($10^{-3}$) & L-BFGS Time (s) \\
\hline
Random & $581.1 \pm 380.2$ & $10.6 \pm 6.2$ & $1073 \pm 206$ \\

Center & $92.5 \pm 94.9$ & $9.5 \pm 5.9$ & $426 \pm 159$ \\

Multitask & $25.4 \pm 24.6$ & $9.0 \pm 5.4$ & $243 \pm 93$ \\

LMC & $38.0 \pm 31.5$ & $9.1 \pm 5.8$ & $302 \pm 127$ \\

Spline & $137.1 \pm 603.8$ & $8.2 \pm 4.9$ & $403 \pm 260$ \\

RBF (cubic) & $39.4 \pm 70.7$ & $8.3 \pm 4.7$ & $318 \pm 217$ \\

RBF (gaussian) & $246.3 \pm 374.7$ & $8.9 \pm 5.0$ & $680 \pm 406$ \\

RBF (multiquadric) & $30.3 \pm 66.9$ & $8.1 \pm 4.6$ & $280 \pm 161$ \\

Polynomial & $13.4 \pm 8.7$ & $8.5 \pm 5.1$ & $249 \pm 129$ \\
\hline
\end{tabular}}
\label{table:diff_performance}
\end{center}

In Table \ref{table:diff_performance} we start to see the interpolation methods come closer to the improvement provided by the center initialization. They both still significantly improve over a randomized initialization. Another trend to note is that in the ''easier" problems, the best methods were the traditional ones; Spline, RBF, and polynomial. Now we observe that Multitask GPs and LMC start to work as well, if not better.

\subsection{6D Parametric Equation}

We consider the following spatially 2D and parametrically 6D diffusion-reaction equation, which is taken from \cite{PSAROS2022111121} which covers metalearning PINN loss functions on parametric PDEs:
\begin{linenomath}\begin{equation}
\left( \frac{\partial^2 u}{\partial x^2} + \frac{\partial^2 u}{\partial y^2}\right) + u \left(1-u^2\right) = f, \, (x,y)\in\Omega
\label{eqn:diffusion-reaction-6d}
\end{equation}\end{linenomath}
\noindent where $\Omega = [-1,1]^2$.  
Here $f$ denotes the source term and we specify an exact solution as $u(x,y;\alpha_1, \alpha_2, \omega_1, \omega_2, \omega_3, \omega_4) = \alpha_1 tanh(\omega_1 x) tanh(\omega_2 y) + \alpha_2 sin(\omega_3 x) sin(\omega_4 y)$ which gives rise to the source term f depending on parameters $(\alpha_1,\alpha_2,\omega_1,\omega_3,\omega_3,\omega_4)$. The bounds for these parameter are $\alpha \in [0.1,1]$ and $\omega \in [1,5]$. We initially randomly sample our parameter space with $17$ PINNs approximations containing $8$ hidden layers of width $20$.

\begin{center}
\captionof{table}{We sample the PDE parameter space with 100 LHS points to compute the values in the table.}
\scalebox{0.75}{
\begin{tabular}{| c | c | c | c |}
\hline
Initialization Methods & Error after Adam ($10^{-3}$) & Error after L-BFGS ($10^{-3}$) & L-BFGS Time (s) \\
\hline
Random & $263.4 \pm 264.3$ & $2.2 \pm 1.3$ & $612 \pm 255$ \\

Center & $90.8 \pm 98.6$ & $1.8 \pm 1.7$ & $494 \pm 222$ \\

Multitask & $73.8 \pm 229.8$ & $1.7 \pm 1.9$ & $431 \pm 200$ \\

LMC & $69.0 \pm 134.7$ & $1.5 \pm 1.8$ & $428 \pm 201$ \\

RBF (multiquadric) & $56.0 \pm 95.9$ & $1.7 \pm 2.0$ & $375 \pm 168$ \\

Polynomial & $109.3 \pm 110.9$ & $1.8 \pm 2.0$ & $496 \pm 213$ \\
\hline
\end{tabular}}
\label{table:6D_performance}
\end{center}

In Table \ref{table:6D_performance} we sub-select to the best performing RBF method and do not perform spline interpolation as it is non-trivial for a 6D problem. The difference between interpolation and center initialization is minimal, but both vastly outperform randomization. Further investigation is necessary into quantifying how different parametric PDEs and domains affect the interpolating methods, particularly at high dimensions.

While we still see improvement with the methods, the amount has decreased compared to the lower dimensional problems. Using more points for the surrogate models has miniminal benefit as shown in Section \ref{ssec:heat} for the Heat problem which was also observed in our experience here. Therefore, there is still work to be done to better apply our method to higher dimensions and grow the improvement in cost and accuracy.

\subsection{10D Parametric Equation}

We consider the following spatially 2D and parametrically 10D equation:
\begin{linenomath}\begin{align}\label{eq:10D_eq}
&- \nabla \cdot [\nabla u (1-\frac{1}{2\pi} \sum_{j=1}^{m} \frac{\bxi^{(j)}}{j} cos(j(x+y)))] = f, \, (x, y) \in \Omega, & \bxi = (\xi^{(1)},...,\xi^{(m)})
\end{align}\end{linenomath}
where $\Omega = [-1,1]^2$, $\bxi = [0,2]^{10}$, and $m$ = 10. We specify an exact solution as $u(x,y;\bxi) = e^{-\sqrt{\bxi^{(1)} x^2+\bxi^{(2)} y^2}} sin(\pi x) + e^{-\sqrt{\bxi^{(3)} x^2+\bxi^{(4)} y^2}} \frac{sin(\pi y)}{4} + e^{-\sqrt{\bxi^{(5)} x^2+\bxi^{(6)} y^2}} \frac{cos(\pi x)}{8} + e^{-\sqrt{\bxi^{(7)} x^2+\bxi^{(8)} y^2}} \frac{cos(\pi y)}{16} + e^{-\sqrt{\bxi^{(9)} x^2+\bxi^{(10)} y^2}} \frac{sin(\pi (x + y))}{32} $ and derive the corresponding form of the forcing $f$. The form of the PDE is analogous to a diffusion equation where the coefficient is the first 10 terms of a Fourier expansion. 

One caveat is that the coefficient term must be positive for $m$, in the case of $m = 10$ and $\bxi = [0,2]^m$, $| 1-\frac{1}{2\pi} \sum_{j=1}^{10} \frac{\bxi^{(j)}}{j} cos(j(x+y))| > 1 - \frac{1}{\pi} \sum_{j=1}^{10} \frac{1}{j} = 0.068 $. The form of the PDE and the exact solution is such that the intrinsic dimension is lowered by reducing the importance of additional PDE parameters. 

\begin{center}
\captionof{table}{We sample the PDE parameter space with 100 LHS points to compute the values in the table.} 
\scalebox{0.75}{
\begin{tabular}{| c | c | c | c |}
\hline
Initialization Methods & Error after Adam ($10^{-3}$) & Error after L-BFGS ($10^{-3}$) & L-BFGS Time (s) \\
\hline
Random & $73.9 \pm 26.3$ & $2.6 \pm 2.0$ & $1762 \pm 516$ \\

Center & $16.9 \pm 11.0$ & $1.7 \pm 0.8$ & $1095 \pm 466$ \\

Multitask & $13.3 \pm 8.4$ & $1.8 \pm 1.0$ & $921 \pm 434$ \\

LMC & $21.0 \pm 17.7$ & $1.6 \pm 0.9$ & $962 \pm 425$ \\

RBF (multiquadric) & $23.7 \pm 42.1$ & $2.1 \pm 2.1$ & $1039 \pm 512$ \\

Polynomial & $12.4 \pm 10.4$ & $1.6 \pm 0.8$ & $973 \pm 468$ \\
\hline
\end{tabular}}
\label{table:10D_performance}
\end{center}


In our observation, the most promising method in terms of cost-benefit is initializing with a run at the center of the parameter domain. However, for the lower-dimensional problems, we do note that interpolating can achieve $10^{-3}$ error without any training, which is very promising. As the dimension and complexity grow, center initialization seems to be the most enticing method. The cost is negligible to do only one run, and in all cases, it is a vast improvement over randomized weights. Center initialization is easily implemented, and we believe it has use whenever working with parameterized PDEs in which one can exploit the similarity in weights in the parametric domain. Further investigation is warranted in regards to identifying task boundaries in the parametric space, so that this method can be used without prior knowledge of the problem. In this paper, we assume constant task regions as elaborated on in Section \ref{sec:parameterPDE}.

\section{Summary and Conclusions}
\label{sec:summary}

In this paper, we have successfully metalearned weight initializations of Physics-informed Neural Networks (PINNs) \cite{raissi2019physics, raissi2017physicsI, raissi2017physicsII} using a model-aware approach on a testbed of parametric PDE problems. Having summarized the metalearning approach, along with the collocating version of PINNs, we gather data using initializations from a fully optimized PINN at the center of the parameter domain. By exploiting the smoothness of parametric PDEs in the weight domain when initialized in this way, we can interpolate optimal weight initialization. Using the data collected, we employ a survey of interpolation methods and empirically show they provide weight initializations that greatly improve optimization. We also compare to the standard model-agnostic metalearning method (MAML)\citep{finn2017model}. These ideas have been successfully implemented in \cite{PENWARDEN2022110844} to speed up multifidelity modeling for PINNs which requires many runs over a parametric domain and we hope it can be directly utilized in other applications as well. Future investigations will explore the problem of task regions and how to approach metalearning for more complex and higher dimensional parametric domains. Along this line of investigation, we also hope to provide a better theoretical understanding of the $\bxi$-variation of weights in a PINN which will help in defining these task regions.

\vspace{0.2in}
\noindent {\bf Acknowledgements:}
The authors would like to acknowledge helpful discussions with Professor George Karniadakis and his group (Brown University).  This work was funded under AFOSR MURI FA9550-20-1-0358.

\appendix 
\section{Symbols and Notations}
\label{sec:appendix}

\begin{center}
\captionof{table}{Symbols and Notations} 
\scalebox{1}{
\begin{tabular}{ ll }
\hline
$u$ & PDE solution \\
$\tilde{u}$ & PINN approximation of PDE solution \\
$m$ & PDE parameter dimension \\
$\mathcal{X}$ & Parameter domain, $\, \mathcal{X} \subset \mathbb{R}^m$ \\
$\bxi$ & Parameter value, $\, \bxi \in \mathcal{X}$ \\
$\bxi^{\mathcal{C}}$ & Centroid of $\mathcal{X}$ \\
$K$ & Number of sample task parameters, $\, \{\bxi_1,...,\bxi_K\} \subset \mathbb{R}^m$ \\ 
$\bxi_i$ & Parameter value in the set, $\, \bxi_i \in \mathcal{X}, \, i = 1,...,K$ \\
$\bxi^{(i)}$ & Parameter value in the i-dimension, $\, \bxi = \left(\xi^{(i)},...,\xi^{(m)}\right)$ \\
$\bXi$ & Matrix of parameter values, $\, \bXi = \left(\bxi_1,...,\bxi_K\right)$ \\
$s$ & Spatial dimension \\
$\Omega$ & Spatial domain of interest, $\, \Omega \subset  \mathbb{R}^p$ \\
$\x$ & Spatial value, $\, \x \in \Omega$ \\
$T$ & Temporal domain of interest \\
$t$ & Temporal value, $\, t \in [0,T]$ \\
$\ell$ & Number of hidden layers \\
$M_{\ell}$ & Number of trainable weights \\
$\underline{w}$ & Trained PINN weight vector \\
$\underline{\hat{w}}$ & Weight predictor, $\, \underline{\hat{w}}: \bxi \rightarrow \underline{w}$ \\
$d$ & Order of hyperbolic cross \\

\hline

\end{tabular}}
\label{table:burgers_performance_largeNN}
\end{center}

\section{Theoretical Considerations}
\label{sec:theoretical}

We provide discussion in this section that motivates why a metalearning ansatz of the form \eqref{eq:linear-approx} can be successful for our core application of approximating solutions to parametric PDEs with PINNs. Our core assumption is that learning the $\bxi$-variation of weights in a PINN is analogous to learning such parametric variation of the solution $u$ itself. We cannot yet make this analogy quantitatively concrete since the behavior of parameters of a trained neural network is still an area of active research. Nevertheless, under this assumption, we analogize the task of learning $\bxi$ variation of neural network weights to that of learning $\bxi$ variation of the PDE solution $u$, and the latter problem is much more well-understood. We emphasize that our discussion below about existing theory for parameter-to-solution maps does \textit{not} yield similar theory for our metalearning framework. However, we believe this discussion is important conceptual motivation for our metalearning strategy.

We first note that for several parametric PDEs of interest, the solution map $\bxi \mapsto u(\cdot, \cdot;\bxi)$ is (real and/or complex) analytic. This fact is known to result in encouraging theoretical guarantees for approximations of parametric PDEs \cite{cohen_approximation_2015}. To be more precise, we furnish two examples. 

First, under the previously described analyticity assumptions, and assuming $\mathcal{X} = [-1,1]^m$, one can conclude the existence of an approximation $u_N$ satisfying,
\begin{linenomath}\begin{align}\label{eq:exponential-rate}
  u_K &\coloneqq \sum_{n=1}^K v_n \phi_n(\bxi), & \left\| u(\bxi) - u_K(\bxi) \right\|_{F(\Omega \times [0,T])} \lesssim \exp\left(-K^{1/m}\right)
\end{align}\end{linenomath}
uniformly for $\bxi \in \mathcal{X}$. Above, $F(\Omega \times [0,T])$ is an appropriate function space over the spacetime variables, the $v_n$ are function-valued coefficients, and $\phi_n$ are \textit{polynomial} functions of $\bxi$. Hence, polynomial approximations can, in principle, yield exponential convergence with efficacy blunted by the curse of dimensionality, as evidenced by the $K^{1/m}$ exponent. We emphasize the parallel between $u_K$ in the expression above and $\hat{\underbar{w}}$ in \eqref{eq:linear-approx}. 

As a second example, one can gain even better rates of convergence, at the cost of making more restrictive technical assumptions. In this case, a similar type of approximation $u_N$ yields \textit{dimension-independent} convergence rates,
\begin{linenomath}\begin{align*}
  \left\| u(\bxi) - u_K(\bxi) \right\|_{F(\Omega \times [0,T])} &\lesssim K^{-r}, & r &= \frac{1}{q} - 1, \hskip 5pt 0 < q < 1,
\end{align*}\end{linenomath}
again, uniformly for $\bxi \in [-1,1]^m$, where $q$ is a constant that depends on the PDE, but not on $x$, $t$, or $\bxi$. Above, the particular $u_N$ approximation differs from the one achieving \eqref{eq:exponential-rate}, but the form of the approximation is identical and in particular utilizes basis functions $\phi_n$ that are still polynomials. The estimate above applies to a particular class of parametric elliptic PDEs under certain assumptions which determine the constant $q$. We refer the reader to \cite[Corollary 7.4]{cohen_convergence_2010} for these assumptions. The relevant message in our context is that $q$ is \textit{independent} of the parameter dimension $m$, demonstrating that it is possible, in principle, to efficiently approximate solutions to parametric PDEs with linear maps of a form similar to \eqref{eq:linear-approx}.

In the PINNs context, our emulator \eqref{eq:linear-approx} does not approximate the solution map itself so that the above theory does not apply. Instead, we build an emulator on the parameters of a neural network. While the behavior of $\bxi$-variability of trained PINNs parameters is not yet fully understood, the theory above suggests that in some situations, one can expect that linear approximations such as \eqref{eq:linear-approx} can perform well in approximating the manifold of well-trained neural network parameters, under the assumption that the trained weights exhibit a similar variation. Our numerical results section explores cases when this indeed is successful, either with polynomials or with other types of approximation procedures. We believe this preliminary analysis and the proof-of-concept of our numerical results can serve as a foundation for future work to better understand the $\bxi$-variation of weights in a PINN.

\section{Methods}
\label{sec:methods}

In this section, we present two categories of methods within the CS\&E world
often applied to parameterized PDE reduced-order modeling:  statistical methods for regression
and numerical methods for approximation.  We only select exemplars from both lists
for our study; we acknowledge that many more methods exist.  However, we hold that these
five methods help give sufficient breadth of application and insight to motivate PINNs metalearning.

\subsection{Statistical Methods for Regression}


Gaussian processes (GPs) are powerful statistic regression models. Due to their nonparametric Bayesian nature~\cite{Rasmussen06GP}, GPs can automatically capture the complexity of the functions underlying the data, and quantify the predictive uncertainty via a closed form (\ie Gaussian). The standard GP learns a single-output stochastic process $f: \mathbb{R}^m\rightarrow \mathbb{R}$ from the training data  $\Dcal = \{(\bxi_1, y_1), \ldots, (\bxi_K, y_K)\}$ where each $\bxi_k$ is an input vector and $\bXi = (\bxi_1,...,\bxi_K)$. The function values $\f = (f(\bxi_1), \ldots, f(\bxi_K))$ are assumed to follow a multivariate Gaussian distribution --- the finite projection of a Gaussian process $\N$ on the training inputs  --- $p(\f| \bXi) = \N(\f|\bmu, \bSigma)$, where $\bmu$ are the mean function values of every input and usually set to $\0$, and the covariance matrix
\begin{linenomath}\begin{align*}
\bSigma = 
\begin{pmatrix} 
\kappa(\bxi_1, \bxi_1)  & \cdots & \kappa(\bxi_1, \bxi_K) \\
\vdots & \ddots & \vdots \\
\kappa(\bxi_K, \bxi_1) & \cdots & \kappa(\bxi_K, \bxi_K)
\end{pmatrix}
\end{align*}\end{linenomath}
where $\kappa$ is a kernel function of the input vectors. The observed outputs $\y$ are assumed to be sampled from a noisy model, \eg $p(\y|\f) = \N(\y|\f, \tau\I)$. Integrating out $\f$, we obtain the marginal likelihood $p(\y|\bXi) = \N(\y|\0, \bSigma + \tau\I)$. Computational estimation of kernel parameters and the noise variance $\tau$ is frequently achieved via maximum likelihood estimation. 

To predict the output for a test input $\bxi^*$, we use the conditional Gaussian distribution, since the test and training outputs jointly follow a multivariate Gaussian distribution. 
\begin{linenomath}\begin{align*}
\textrm{We have } & \, p\big(f(\bxi^*)|\bxi^*, \bXi,\y\big) = \N\big(f^* | \alpha(\bxi^*), \beta(\bxi^*)\big) \\ \textrm{where } & \, \alpha(\bxi^*) = \kappa_{*}^\top (\bSigma + \tau^{-1} \I)^{-1}\y, \\ & \beta(\x^*) = \kappa(\bxi^*, \bxi^*) - \kappa_{*}^\top(\bSigma + \tau^{-1}\I)^{-1}\kappa_{*}, \\ & \, \kappa_{*} = [\kappa(\bxi^*, \bxi_1), \ldots, \kappa(\bxi^*, \bxi_K)]^\top.
\end{align*}\end{linenomath}
In particular, we use the mean of $p\big(f(\bxi^*)|\bxi^*, \bXi,\y\big)$ as the predictor at $\bxi^*$. These methods are implemented with GPyTorch \citep{gpytorch}.

\subsubsection{Multi-task Gaussian Process Modeling}
Many tasks  require learning a function with multiple output.\footnote{Task here is used differently than in the metalearning section. We use the nomenclature of the GP field here.} A classical multi-output regression framework is multi-task GP~\cite{williams2007multi}. It models the outputs of all the tasks as a single GP, where the kernel function between arbitrary two outputs is defined as 
\begin{linenomath}\begin{align}
	g([\bxi_1, t_1], [\bxi_2, t_2]) = \kappa(\bxi_1, \bxi_2) \cdot s_{t_1 t_2},
\end{align}\end{linenomath}
where $\bxi_1, \bxi_2$ are the inputs and $t_1, t_2$ are the task indices, $\kappa(\cdot, \cdot)$ is a kernel function of the inputs, and $s_{t_1t_2}$ is the task similarity between $t_1$ and $t_2$. The training is the same as the standard GP. However, we need to estimate not only the parameters of $\kappa(\cdot, \cdot)$, but also the similarity values between each pair of tasks $\{s_{t_1t_2}\}$.

\subsubsection{Linear Method of Coregularization (LMC)}


Another successful multi-output regression model is  the Linear Model of Coregionalization (LMC)~\cite{journel1978mining}, which assumes the outputs are a linear combination of a set of basis vectors weighted by independent random functions. We introduce $Q$ bases $\B = [\b_1, \ldots, \b_Q]^\top$ and model a $M_{\ell}$-dimensional vector function by 
\begin{linenomath}\begin{align}
	\hat{\underline{w}}(\bxi) = \sum_{q=1}^{Q} \b_q c_q(\bxi) = \B^\top \c(\bxi) \label{eq:LMC}
\end{align}\end{linenomath}
where $Q$ is often chosen to be much smaller than $M_{\ell}$, and the random weight functions $\c(\bxi) = [c_1(\bxi), \ldots, c_Q(\bxi)]^\top$ are sampled from independent GPs. 
In spite of a linear structure, the outputs $\hat{\underline{w}}$ are still nonlinear to the input $\bxi$ due to the nonlinearity of the weight functions. LMC can easily scale up to a large number of outputs: once the bases $\B$ are identified, we only need to estimate a small number of GP models ($Q \ll M_{\ell}$). For example, we can perform PCA on the training outputs to find the bases, and use the singular values as the outputs to train the weight functions. This is also referred to as PCA-GP~\cite{higdon2008computer}. 

\subsection{Numerical Approximation Methods for Regression}

We describe here common strategies in the numerical approximation community for building emulators of the form \eqref{eq:linear-approx} from data. We choose to focus on three different types of approaches: First we build approximations via cubic spline interpolants when $\bxi$ is one or two dimensional. The second class of methods use ``meshless" radial basis function (RBF) methods with various kernels. The third class of methods are polynomial approximation schemes using least squares approximation.

We briefly describe these approaches in each section below, focusing on specifying how the coefficients $\b_k$ and basis functions $c_k(\bxi)$ in \eqref{eq:linear-approx} are chosen and computed. In all the methods, we assume availability of data $\underbar{w}(\bxi_j)$ (i.e., trained PINNs network parameters) on a discrete sampling $\{\bxi_i\}_{i=1}^{K}$. We will specify how this grid is chosen in each section.

\subsubsection{Cubic Spline Interpolation}
\label{ssec:spline}
Univariate cubic splines are smooth interpolants of data \cite{boor_practical_1978}. More precisely, with $\left(\bxi_i\right)_{i=1}^{K}$ an ordered set of points from a Latin Hypercube Sampling (LHS) design \cite{lemieux_monte_2009} on the one-dimensional interval $\mathcal{X}$, then a cubic spline approach builds the approximation $\hat{\underbar{w}}$ in \eqref{eq:linear-approx} as a piecewise cubic function, where $\hat{\underbar{w}}$ is a cubic polynomial on each interval $[\bxi_j, \bxi_{j+1}]$, and is continuously differentiable at each $\bxi_j$. One can build such an approximation in terms of divided differences of the data $\underbar{w}(\bxi_j)$, so that the coefficients $\b_k$ are linear in this data. The basis functions $c_k(\bxi)$ are then piecewise cubic polynomials. This is implemented with the SciPy package functions scipy.interpolate.UniveriateSpline and scipy.interpolate.SmoothBivariateSpline \citep{2020SciPy-NMeth}. The bivariate case does not require a tensoral grid therefore we can use it with our sampling methods with the implemtation caveat that $K \geq (x_{deg}+1)*(y_{deg}+1)$ which in the cubic spline case is $K \geq 16$. Additionally, we use the default function parameters.


\subsubsection{Radial Basis Function (RBF) Interpolation}

RBF approximations build the emulator $\hat{\underbar{w}}$ in \eqref{eq:linear-approx} using shifted kernels as basis functions,
\begin{linenomath}\begin{align*}
  c_k(\bxi) = \kappa(\bxi, \bxi_k),
\end{align*}\end{linenomath}
and the coefficients $\b_k$ are chosen by enforcing interpolation of $\hat{\underbar{w}}$ on the data points \cite{fasshauer_meshfree_2007}. We again choose the parametric grid as an LHS design, and employ three different types of commonly used kernels:
\begin{linenomath}\begin{align*}
  \kappa(\bxi, \boldsymbol{\zeta}) &= r^3, & \textrm{(Cubic)} \\
  \kappa(\bxi, \boldsymbol{\zeta}) &= \exp\left(-\left(\frac{r}{\tau}\right)^2\right), & \textrm{(Gaussian)}\\
  \kappa(\bxi, \boldsymbol{\zeta}) &= \sqrt{\left(\frac{r}{\tau}\right)^2 + 1} , & \textrm{(Multiquadric)}
\end{align*}\end{linenomath}
where $r = \left\| \bxi - \boldsymbol{\zeta} \right\|_2$ is the Euclidean distance between the inputs $\bxi$ and $\boldsymbol{\zeta}$, and $\tau$ is a tunable parameter, which we choose as the average pairwise distance between grid points. 

We use interpolative RBF approximations, which enforce \eqref{eq:linear-approx} at every data point. Such RBF approximations are closely related to GP statistical models: an RBF approximation equals the mean of a GP approximation built from the same kernel, with zero nugget at the data points.

\subsubsection{Polynomial least-squares using hyperbolic cross sets}
\label{sec:cross-sets}

Our final technique in this section for forming the approximation \eqref{eq:linear-approx} builds a polynomial approximation via least squares. In this case, the basis functions are polynomials,
\begin{linenomath}\begin{align*}
  c_k(\bxi) &= \bxi^{\alpha(k)} = \prod_{j=1}^m \left(\xi^{(j)}\right)^{\alpha(k)_j}, & \bxi &= \left( \xi^{(1)}, \ldots, \xi^{(m)}\right),
\end{align*}\end{linenomath}
where $\alpha(k) = (\alpha(k)_1, \ldots, \alpha(k)_m)\in \N_0^{m}$ is an $m$-dimensional multi-index on the non-negative integer lattice, and $\left(\alpha(k)\right)_{k=1}^K$ is any ordering of the multi-indices lying in an order-$d$ \textit{hyperbolic cross} set \cite{dung_hyperbolic_2018},
\begin{linenomath}\begin{align*}
  \left(\alpha(k)\right)_{k=1}^K \eqqcolon A = \left\{ \alpha \in \N_0^m \;\big|\; \prod_{j=1}^m (\alpha_j + 1) \leq d + 1 \right\}.
\end{align*}\end{linenomath}
Increasing the order $d$ increases the number of terms $K$ in the expansion, and hyperbolic cross sets tend to de-emphasize mixed polynomial terms and therefore have much smaller size than alternative total degree sets. Despite this decreased model capacity, polynomial approximations on hyperbolic cross sets are known to produce efficient approximations in the context of parametric PDEs \cite{dung_hyperbolic_2016,chkifa_polynomial_2018}.

We construct the coefficients $\b_k$ in \eqref{eq:linear-approx} using a weighted least squares approach on the data, and the grid is chosen as a set of weighted approximate Fekete points \cite{guo_weighted_2018}. This method is implemented with the UncertainSCI package \citep{uncertainsci}.

\section{Additional computational results}
\label{sec:additional}

To demonstrate the use of the center of the domain over other arbitrary points, we provide the following results. We run experiments for the problems in Sections \ref{ssec:burgers} and \ref{ssec:heat} three additional times, where a random point in the parametric domain is used instead of the center value. This is done for the single parameter case of Burgers and the double parameter case in the Heat problem. The bounds for these parameters are $[0.005,0.05]$ and $[1,\pi]\times[1,\pi]$ respectively. The results for (Random) and (Center) were provided for easy comparison and are the same as in Table \ref{table:burgers_performance} and Table \ref{table:heat_performance} of the manuscript. Recall, (Random) in the manuscript tables refers to random initialization of the weights, not an arbitrary random point in the domain instead of the center which is being discussed here.

\begin{center}
\captionof{table}{Random parameter runs used as initialization instead of the center of the parametric domain}
\scalebox{0.75}{
\begin{tabular}{| c | c | c | c | c |}
\hline
PDE & Parameter Value & Error after Adam ($10^{-3}$) & Error after L-BFGS ($10^{-3}$) & L-BFGS Time (s) \\
\hline
Burgers (Random)& $None$ & $618.1 \pm 177.7$ & $5.1 \pm 6.0$ & $146 \pm 68$ \\

Burgers (Center)& $0.0275$ & $58.6 \pm 103.7$ & $6.4 \pm 9.1$ & $63 \pm 51$ \\

Burgers 1 & $0.00769$ & $170.4 \pm 98.9$ & $1.8 \pm 1.5$ & $90 \pm 38$ \\

Burgers 2 & $0.04055$ & $54.8 \pm 88.3 $ & $3.8 \pm 3.9$ & $80 \pm 64$ \\

Burgers 3 & $0.01729$ & $38.6 \pm 53.5$ & $2.9 \pm 4.6$ & $59 \pm 37$ \\

\hline

Heat (Random)& $None$ & $465.1 \pm 218.2$ & $5.5 \pm 3.7$ & $178 \pm 37$ \\

Heat (Center)& $[2.07080, 2.07080]$ & $38.4 \pm 28.8$ & $4.6 \pm 2.8$ & $102 \pm 40$ \\

Heat 1 & $[2.39901, 1.99476]$ & $95.8 \pm 55.0$ & $3.7 \pm 2.5$ & $157 \pm 53$ \\

Heat 2 & $[1.05285, 1.14402]$ & $295.8 \pm 24.8$ & $7.5 \pm 5.7$ & $173 \pm 63$ \\

Heat 3 & $[1.09170, 2.59512]$ & $71.0 \pm 60.3$ & $7.1 \pm 5.4$ & $128 \pm 44$ \\

\hline
\end{tabular}}
\label{table:off-center_runs}
\end{center}

In the table we can see two trends between both PDE problems. First is that the center initialization has the smallest or close to the smallest L-BFGS training time. Burgers 3 runtime is slightly smaller than Burgers (Center) however this value is relatively close to the center value. Next we observe that the value drawn close to the bounds had significantly higher error after Adam optimization and the longest L-BFGS runtimes of the values besides random initialization. For Burgers, this is seen at a low viscosity meaning the shock is more sharp and the problem more difficult. This makes sense as the solution is initially more different than most cases in the parametric domain compared to a run at the center, so it takes more training to transition to the other parametric solutions than something which starts at the center. In regard to final error after L-BFGS, the results are conflicting, the lower viscosity run has the lowest final error, perhaps because it is more suited to the difficult region of the domain. However, in the Heat problem the run close to both lower bounds (Heat 2) has the largest final error. Keeping in mind there is also inherently variation when training neural networks the effect on final error is not conclusive. What we do find is conclusive is that using the center value is a reasonable setting. Potentially, a more optimal single value does exist but it is not clear how that would be discovered apriori. Rather, we state this is where the benefit of our surrogate modeling methods come in to find a more optimal initialization over the entire domain at the cost of pre-training.

\newpage
\noindent {\bf Bibliography}

\bibliography{references,metalearning-ref}
\end{document}